\documentclass[final,3p,times]{elsarticle}

\usepackage{amssymb}

\usepackage{amssymb}
\usepackage[]{amsmath}
\usepackage{graphicx}
\usepackage{graphics}
\usepackage{subfigure}
\usepackage{latexsym}
\usepackage[]{amsmath}
\usepackage{amssymb}
\usepackage{indentfirst}
\usepackage{epsfig}
\usepackage{epstopdf}
\usepackage{float}
\usepackage{setspace}
\usepackage{booktabs}
\usepackage{times}
\usepackage{anysize}

\biboptions{numbers,sort&compress}
\usepackage[colorlinks, citecolor=red]{hyperref}
\usepackage{amsmath}
\usepackage{times}
\usepackage{anysize}
\marginsize{1.5cm}{1.5cm}{1.2cm}{1.5cm}
\selectfont

 \journal {}

\begin{document}

\begin{frontmatter}



\title{Nonlocal symmetries and exact solutions of the (2+1)-dimensional generalised variable coefficient shallow water wave equation}

\author{ Xiangpeng Xin \fnref{label2} $^*$  }

\cortext[cor1]{Corresponding author}
\ead{xinxiangpeng2012@gmail.com}

\author{Linlin Zhang \fnref{label2} }
\author{Yarong Xia \fnref{label3} }
\author{Hanze Liu \fnref{label2} }
\address[label2]{ School of Mathematical Sciences, Liaocheng University, Liaocheng 252059, PR China}
\address[label3]{ School of Information and Engineering, Xi'an University, Xi'an 710065, PR China}

\begin{abstract}
In this paper, using the standard truncated Painlev\'{e} analysis, the Schwartzian equation of (2+1)-dimensional generalised variable coefficient shallow water wave (SWW)equation is obtained. With the help of lax pairs, nonlocal symmetries of the SWW equation are constructed which be localized by a complicated calculation process. Furthermore, using the Lie point symmetries of the closed system and Schwartzian equation, some exact interaction solutions are obtained, such as soliton-cnoidal wave solutions. Corresponding 2D and 3D figures are placed on illustrate dynamic behavior of the generalised variable coefficient SWW equation.

\end{abstract}

\begin{keyword}
Truncated Painlev\'{e} analysis; Nonlocal symmetry; Interaction solution.

\end{keyword}
\end{frontmatter}


\section{Introduction}

Searching for new exact solutions to nonlinear partial differential equations(NPDEs) which appear in many branches of nonlinear science is always an interesting topic. Many methods have been proposed to address the NPDEs, such as the bilinear method, Painlev\'{e} analysis, symmetry reductions and B\"{a}cklund and Darboux transformations etc.

As we all know, Lie group analysis\cite{Olver1,Bluman1,Bluman2,Liu1} and Painlev\'{e} analysis\cite{W,C1,P1,Liu2} are very effective methods for constructing exact and analytic solutions of NPDEs. Recently, the nonlocal symmetry method\cite{Bluman4,Lou3} which related to the infinitesimal forms of lax pairs or B\"{a}cklund transformations has been widely applied to construct interaction solutions of some nonlinear systems. Lou et al.\cite{Hu2,Xin1,H1,Xin2,Ch1,Hu} have pointed out that there is a close relationship between Painlev\'{e} analysis and nonlocal symmetries, i.e. the last variable introduced in the localization process of nonlocal symmetry must satisfy the Schwartzian form of the equation, and this theory greatly simplifies the process of constructing exact solutions to closed systems. Nowadays, Lie group theory has been widely applied to many fields, such as, mathematics, physics, numerical analysis\cite{Xia1,Xia2}, quantum mechanics, fluid dynamics system, etc.

On the other hand, constructing nonlocal symmetries and exact solutions of variable coefficients NPDEs\cite{Miao1} is much more difficult than constant coefficient NPDEs. It is more meaningful to construct the nonlocal symmetric and exact solutions of the variable coefficient equation since the constant coefficient equation is a special form of variable coefficient equation. In this paper, we would like to search for exact interaction solution of (2+1)-dimensional variable coefficient shallow water wave equation using a systemic method.

This paper is arranged as follows: In Sec.2, from the truncated Painlev\'{e} expansion, the Schwartzian form of the (2+1)-dimensional generalised variable coefficient SWW equation is obtained. In Sec.3, the nonlocal symmetries are constructed by using lax pairs of the generalised variable coefficient SWW equation. In Sec.4, the process of transforming from nonlocal symmetries to local symmetries is introduced in detail and the nonlocal symmetry group is found by solving the initial value problem. Symmetry reductions and exact interaction solutions to the SWW are obtained by using the Lie point symmetry of the closed system in Sec.5. Finally, some conclusions and discussions are provided in Sec.6.

\section{Truncated Painlev\'{e} analysis of generalised variable coefficient SWW equation}

In this article, we will investigate the following (2+1)-dimensional generalised variable-coefficient SWW equation\cite{Lan},

 \begin{equation}\label{eq1}
u_{xt}  + 2au_x u_{xy}  + au_y u_{xx}  + bu_{xy}  + \frac{1}{2}\rho au_{xxxy}  = 0,
 \end{equation}
where $(x, y)$ are space coordinate and $t$ is time coordinates, $u(x,y,t)$ is the amplitude of the Riemann wave, $a(t), b(t)$ are real functions of $t$, and $\rho$ is an arbitrary constant. When the functions $a(t), b(t)$  are given special values, the equation(\ref{eq1}) degenerates into usual NPDEs. For example, when $a(t)=-2$, $b(t)=constant \neq0$, the (\ref{eq1}) becomes a (2+1)-dimensional extended shallow water wave equation. when $a(t)=-2$, $b(t)=0$, the (\ref{eq1}) becomes the (2+1)-dimensional breaking soliton equation etc. Before study nonlocal symmetries of Eq.(\ref{eq1}), let's construct the Schwartzian form of this equation firstly.

For the Eq.(\ref{eq1}), we take a truncated Painlev\'{e} expansion. Balancing the nonlinear and dispersive terms, we have the truncated Painlev\'{e} expansion in the form,

 \begin{equation}\label{eq2}
u = \frac{{u_0 }}{\phi } + u_1 ,
 \end{equation}
where $u_0,u_1$ and $\phi$ are arbitrary functions with respect to $x, y$ and $t$. Substituting Eq.(\ref{eq2}) into Eq.(\ref{eq1}) and vanishing coefficients of the different powers  $\frac{{1 }}{\phi }$, we obtain,

\begin{center}
$u_0  = 2\phi _x$, $u_1  =  - \frac{1}{2}\frac{{\phi _{xx} }}{{\phi _x }} - \frac{1}{4}\phi _x \int {\frac{{\phi _{xx}^2 }}{{\phi _x^2 }}dx}$,
\end{center}
which yields the solution of Eq.(\ref{eq1}) as follows:

\begin{equation}\label{eq3}
u = \frac{{2\phi _x }}{\phi } - \frac{1}{2}\frac{{\phi _{xx} }}{{\phi _x }} - \frac{1}{4}\phi _x \int {\frac{{\phi _{xx}^2 }}{{\phi _x^2 }}dx},
\end{equation}
and Eq.(\ref{eq1})is successfully reduced to its Schwarzian form,
\begin{equation}\label{eq4}
C_x  + a SK_x  + \frac{1}{2}a KS_x  + \frac{1}{2}a K_{xxx}  + bK_x  = 0,
\end{equation}
with $C = \frac{{\phi _t }}{{\phi _x }},K = \frac{{\phi _y }}{{\phi _x }},S = \frac{{\phi _{xxx} }}{{\phi _x }} - \frac{3}{2}\frac{{\phi _{xx}^2 }}{{\phi _x^2 }}.$ The Schwarzian equation (\ref{eq4}) is invariant under the M\"{o}bious transformation,
\begin{center}
$\phi  \to \frac{{A_1  + A_2 \phi }}{{A_3  + A_4 \phi }},~~~~~A_1 A_4  \ne A_2 A_3$.
\end{center}

From standard truncated Painlev\'{e} expansion Eq.(\ref{eq1}), we have the above non-auto-B\"{a}cklund transformation(\ref{eq3}) i.e., if the function $\phi$ satisfies Eq.(\ref{eq4}), then (\ref{eq3})is just the solution of the (2+1)-dimensional generalised variable coefficient SWW equation(\ref{eq1}).

\section{Nonlocal symmetries of generalised variable coefficient SWW equation}

To our knowledge, nonlocal symmetries for Eq.(\ref{eq1}) have not been obtained and discussed, which will be the goal of this paper. The corresponding Lax pairs has been obtained in\cite{Lan},

 \begin{equation}\label{eq5}
\begin{array}{l}
 \psi _{xx}  =  - \frac{1}{\rho }u_x \psi , \\
 \psi _t  =  - \frac{1}{2}\rho a\psi _{xxy}  - a\psi _x u_y  - \frac{1}{2}a\psi _y u_x  - b\psi _y , \\
 \end{array}
 \end{equation}
with the equivalent compatibility condition $\psi _{xxt}=\psi _{txx}$.

The next work is to construct the nonlocal symmetries of generalised variable coefficient SWW equation(\ref{eq1}). Known from the lie group theory, the symmetries of $u,a,b$ have to satisfy the following linearized equation,
 \begin{equation}\label{eq6}
\begin{array}{l}
 2\sigma ^2 u_x u_{xy}  + 2a\sigma _{xy}^1  + 2a\sigma _x^1 u_{xy}  + \sigma ^2 u_{xx} u_y  + a\sigma _{xx}^1 u_y  \\
  + a\sigma _y^1 u_{xx}  + \frac{1}{2}\rho \sigma ^2 u_{xxxy}  + \frac{1}{2}\rho a\sigma _{xxxy}^1  + \sigma _{xt}^1  + \sigma ^3 u_{xy}  + b\sigma _{xy}^1  = 0, \\
 \end{array}
 \end{equation}
where $\sigma ^1,\sigma ^2,\sigma ^3$ are symmetries of $u,a$ and $b$, which means Eq.(\ref{eq1}) is form invariant under the following infinitesimal transformations,
\begin{center}
$\begin{array}{l}
 u \to u + \epsilon \sigma ^1 , \\
 a \to a + \epsilon \sigma ^2 , \\
 b \to b + \epsilon \sigma ^3 , \\
 \end{array}$
\end{center}
with the infinitesimal parameter $\epsilon$. Assuming symmetries $\sigma ^1,\sigma ^2,\sigma ^3$ have the following form,

 \begin{equation}\label{eq7}
\begin{array}{l}
 \sigma ^1  = \bar Xu_x  + \bar Yu_y  + \bar Tu_t  - \bar U, \\
 \sigma ^2  = \bar Ta_t  - \bar B_1 , \\
 \sigma ^3  = \bar Tb_t  - \bar B_2 , \\
 \end{array}
 \end{equation}
where $\bar X,\bar Y,\bar T,\bar U,\bar B_1,\bar B_2$ are functions of$(x,y,t,u,a,b,\psi ,\psi _x ,\psi _y )$.

In order to solve variables $\bar X,\bar Y,\bar T,\bar U,\bar B_1,\bar B_2$, we substitute Eq.(\ref{eq7}) into Eq.(\ref{eq6}) and eliminate $u_{xt}, \psi_{xx}, \psi_{t}$ , then obtain a system of determining equations for variables $\bar X,\bar Y,\bar T,\bar U,\bar B_1,\bar B_2$ , solving these determining equations yields,

 \begin{equation}\label{eq8}
\begin{array}{l}
 \bar X = c_1 x + F_3 (t),\bar Y = F_2 (y,t),\bar T = F_1 (t), \\
 \bar U =  - c_1 u + c_2 x + c_3 \psi ^2  + \frac{y}{a}F_{3t} (t) + F_4 (t,a,b), \\
 \bar B_1  = a(2c_1  - F_{1t} (t) + F_{2y} (y,t)), \\
 \bar B_2  = bF_{2y} (y,t) + F_{2t} (y,t) - bF_{1t} (t) - 2c_2 a, \\
 \end{array}
 \end{equation}
where $c_i(i = 1, ..., 3)$ are three arbitrary constants and $F_j(j = 1, ..., 4)$ are arbitrary functions of the corresponding variables. It is shown that the results(\ref{eq8}) are nonlocal symmetries of generalised variable coefficient SWW equation because $\bar U$ is a function of $\psi$. We know that nonlocal symmetries need to be transformed into local symmetries before construct exact solutions. Hence, we need to construct a closed system whose Lie symmetries contain above nonlocal symmetries.

\section{Localization of the nonlocal symmetry}

For simplicity, letting $F_1 (t) = 0,F_2 (y,t) = 0,F_3 (t) = 0,F_4 (t,a,b) = 0,c_1  = 0,c_2  = 0,c_3  =  - 1$ and substituting Eq.(\ref{eq8})into Eq.(\ref{eq7})yields,

\begin{equation}\label{eq81}
 \sigma ^1  = \psi ^2 ,\sigma ^2  = 0,\sigma ^3  = 0.
\end{equation}

To localize the nonlocal symmetry (\ref{eq81}), one should to solve the following linearized equations of Eqs.(\ref{eq5}),

 \begin{equation}\label{eq9}
\begin{array}{l}
 \sigma _{xx}^4  + \frac{1}{\rho }\psi \sigma _x^1  + \frac{1}{\rho }\sigma ^4 u_x  = 0, \\
 \frac{1}{2}\rho \sigma ^2 \psi _{xxy}  + \frac{1}{2}\rho a\sigma _{xxy}^4  + \sigma ^2 u_y \psi _x  + a\sigma _y^1 \psi _x  + a\sigma _x^4 u_y  \\
  + \frac{1}{2}au_x \sigma _y^4  + \frac{1}{2}\psi _y \sigma ^2 u_x  + \frac{1}{2}a\psi _y \sigma _x^1  + b\sigma _y^4  + \psi _y \sigma ^3  + \sigma _t^4  = 0, \\
 \end{array}
 \end{equation}
which is form invariant under the infinitesimal transformation $\psi  \to \psi  + \epsilon \sigma ^4$, where $\sigma ^1,\sigma ^2,\sigma ^3$ given by (\ref{eq81}). One can verify that Eqs.(\ref{eq9}) have the following special solution,

\begin{equation}\label{eq91}
\sigma ^4  = \psi \phi,
\end{equation}
where $\phi$ satisfies the following compatible equations,

 \begin{equation}\label{eq10}
\begin{array}{l}
 \phi _x  =  - \frac{{\psi ^2 }}{{2\rho }}, \\
 \phi _t  = \frac{{au_y \psi ^2  - 2\rho a\psi _x \psi _y  + 2\rho a\psi \psi _{xy}  - 2\rho b\psi \phi _y }}{{2\rho }}. \\
 \end{array}
 \end{equation}

It is not difficult to verify that the auxiliary dependent variable $\phi$ just satisfies the Schwartzian form (\ref{eq4})of the variable coefficient SWW equation and
\begin{equation}\label{eq101}
\sigma ^5=\phi^2,
\end{equation}
which is form invariant under the transformation $\phi  \to \phi  + \epsilon \sigma ^5$. Accordingly, the nonlocal symmetry (\ref{eq81}) has been successfully localized to a Lie point symmetry in the enlarged system. Correspondingly, the initial value problem of the closed system becomes,

\begin{equation}\label{eq102}
\begin{array}{l}
 \frac{{d\bar u(\epsilon )}}{{d\epsilon }} = \psi ^2 ,~~~~~~~~~~~~\bar u(\epsilon )\left| {_{\epsilon  = 0}  = u,} \right. \\
 \frac{{d\bar a(\epsilon )}}{{d\epsilon }} = 0,~~~~~~~~~~~~~~\bar a(\epsilon )\left| {_{\epsilon  = 0}  = a,} \right. \\
 \frac{{d\bar b(\epsilon )}}{{d\epsilon }} = 0,~~~~~~~~~~~~~~\bar b(\epsilon )\left| {_{\epsilon  = 0}  = b,} \right. \\
 \frac{{d\bar \psi (\epsilon )}}{{d\epsilon }} = \psi \phi ,~~~~~~~~~~\bar \psi (\epsilon )\left| {_{\epsilon  = 0}  = \psi ,} \right. \\
 \frac{{d\bar \phi (\epsilon )}}{{d\epsilon }} = \phi ^2 ,~~~~~~~~~~\bar \phi (\epsilon )\left| {_{\epsilon  = 0}  = \phi ,} \right. \\
 \end{array}
\end{equation}
where $\epsilon$ is the group parameter.The solution of Eqs.(\ref{eq102}) leads to the following group theorem for the enlarged system.

Theorem 1. If $\{u, a,b,\psi,\phi\}$ is the solution of the prolonged system (\ref{eq1}),(\ref{eq5})and (\ref{eq10}), so is $\{\bar u, \bar a,\bar b,\bar \psi,\bar \phi\}$
\begin{equation}\label{eq103}
\bar u = \frac{{\epsilon \phi u - \epsilon \psi ^2  + u}}{{\epsilon \phi  + 1}},~~\bar a = a,~~\bar b = b,~~\bar \psi  = \frac{\psi }{{\epsilon \phi  + 1}},~~\bar \phi  = \frac{\phi }{{\epsilon \phi  + 1}},
\end{equation}
with $\epsilon$  is an arbitrary group parameter. From the above theorem, we know that one can get new forms of $u$ from the old solutions of Eqs.(\ref{eq1}),(\ref{eq5})and (\ref{eq10}). For example, we just take a simple solution of Eq.(\ref{eq1}) as,
\begin{center}
$u = 2\rho \tanh (x),$
\end{center}
then $\psi$ and $\phi$ have the following solutions,
\begin{center}
$\begin{array}{l}
 \psi  = \frac{{\cosh (2x)\ln (\cosh (2x) + \sinh (2x)) - 2\sinh (2x) - \ln (\cosh (2x) + \sinh (2x))}}{{\sinh (2x)}}, \\
 \phi  = \frac{{x(x + 2)\ln (e^{2x} )}}{\rho } - \frac{{(xe^{2x}  + x + 2)\ln (e^{2x} )^2 }}{{2\rho (e^{2x}  + 1)}} - \frac{{2x(x^2  + 3x + 3)}}{{3\rho }}, \\
 \end{array}$
\end{center}
then, new exact solution of the equation(\ref{eq1}) is obtained by substituting above $u,\psi,\phi$ into(\ref{eq103}).

To search for more similarity reductions and exact solutions of Eq.(1), we use classical Lie group method to the prolonged system (\ref{eq1}),(\ref{eq5})and (\ref{eq10}),and assuming symmetries have the following form,

\begin{equation}\label{eq104}
\begin{array}{l}
 \sigma ^1  = Xu_x  + Yu_y  + Tu_t  - U, \\
 \sigma ^2  = Ta_t  - B_1 , \\
 \sigma ^3  = Tb_t  - B_2 , \\
 \sigma ^4  = X\psi _x  + Y\psi _y  + T\psi _t  - P_1 , \\
 \sigma ^5  = X\phi _x  + Y\phi _y  + T\phi _t  - P_2 , \\
 \end{array}
\end{equation}
where $X,Y,T,U,B_1,B_2,P_1,P_2$ are functions of$(x,y,t,u,a,b,\psi ,\phi)$. The symmetryis $\sigma^i, (i = 1, ..., 4)$ satisfy the linearized equations (\ref{eq6}),(\ref{eq9})and
 \begin{equation}\label{eq11}
\begin{array}{l}
 \sigma _x^5  + \frac{1}{\rho }\sigma ^4 \psi  = 0, \\
 \sigma ^3 \phi _y  + b\sigma _y^5  + \sigma _t^5  + \sigma ^4 \psi _y \psi _x  + a\sigma _x^4 \psi _y  - \sigma ^2 \psi \psi _{xy}  \\
  - a\sigma ^4 \psi _{xy}  + a\sigma _y^4 \psi _x  - \frac{{\sigma ^2 \psi ^2 u_y }}{{2\rho }} - \frac{{a\psi ^2 \sigma _y^1 }}{{2\rho }} - \frac{{a\sigma ^4 \psi u_y }}{\rho } = 0. \\
 \end{array}
 \end{equation}

Substituting Eqs.(\ref{eq104}) into Eqs.(\ref{eq6},\ref{eq9},\ref{eq11}), and eliminating $u_{x,t},\psi_{xx},\psi_t,\phi_{x},\phi_t$ in terms of the prolonged system, determining equations for
the variables $X,Y,T,U,B_1,B_2,P_1,P_2$ can be obtained, by calculating one can get,

\begin{equation}\label{eq12}
\begin{array}{l}
 X = c_1 x + F_3 (t),Y = F_2 (y,t),T = F_1 (t),U =  - c_1 u + c_2 \psi ^2  + \frac{y}{a}F_{3t} (t) + F_4 (t,a,b), \\
 B_1  = a(2c_1  - F_{1t} (t) + F_{2y} (y,t)),B_2  = bF_{2y} (y,t) + F_{2t} (y,t) - bF_{1t} (t) - 2c_2 a, \\
 P_1  = (c_2 \phi  + c_3 )\psi ,P_2  = c_2 \phi ^2  + (c_1  + 2c_3 )\phi  + c_4 , \\
 \end{array}
\end{equation}
where $c_i(i = 1, ..., 4)$ are three arbitrary constants and $F_j(j = 1, ..., 4)$ are arbitrary functions of the corresponding variables.

\section {Symmetry reduction and exact solutions of variable coefficient SWW equation}

In this section, we will give a nontrivial similarity reduction by using the lie group method. corresponding group invariant solutions of variable coefficient SWW equation(\ref{eq1})are obtained. Without loss of generality, we let $c_1  = 0,c_2  =  - \alpha ,c_3  = 0,c_4  = c_4 ,F_1 (t) = m\beta ,F_2 (y,t) = n\beta ,F_3 (t) = \beta$. By solving the following characteristic equations,

\begin{equation}\label{eq13}
\frac{{dx}}{\beta } = \frac{{dy}}{{n\beta }} = \frac{{dt}}{{m\beta }} = \frac{{du}}{{ - \alpha \psi ^2 }} = \frac{{da}}{0} = \frac{{db}}{0} = \frac{{d\psi }}{{ - \alpha \psi \phi }} = \frac{{d\phi }}{{c_4  - \alpha \phi ^2 }},
\end{equation}
one can obtain,
\begin{equation}\label{eq14}
\begin{array}{l}
 a = \tilde a,b = \tilde b,\phi  = \frac{{\sqrt {c_4 \alpha } \tanh (\frac{{\sqrt {c_4 \alpha } (t + \Phi )}}{{m\beta }})}}{\alpha }, \\
 \psi  = \Psi \sqrt {\tanh ^2 (\frac{{\sqrt {c_4 \alpha } (t + \Phi )}}{{m\beta }} - 1} , \\
 u = \frac{{\alpha \Psi ^2 \tanh (\frac{{\sqrt {c_4 \alpha } (t + \Phi )}}{{m\beta }} + \sqrt {c_4 \alpha } U}}{{\sqrt {c_4 \alpha } }}, \\
 \end{array}
\end{equation}
where $\Phi  = \Phi (\xi ,\eta ),\Psi  = \Psi (\xi ,\eta ),U = U(\xi ,\eta ),\xi  = \frac{{mx - t}}{m},\eta  = \frac{{my - nt}}{m},$ and $\tilde a,\tilde b$ are arbitrary constant.

Substituting Eqs.(\ref{eq14})into the prolonged system yields,
\begin{equation}\label{eq15}
 \begin{array}{l}
 \Psi  =  \pm \frac{{\sqrt {2m\beta \rho c_4 \Phi _\xi  } }}{{m\beta }}, \\
 U = \int {\frac{{\rho (m^2 \beta ^2 \Phi _{\xi \xi }^2  - 4c_4 \alpha \Phi _\xi ^4  - 2m^2 \beta ^2 \Phi _\xi  \Phi _{\xi \xi \xi } )}}{{4m^2 \beta ^2 \Phi _\xi ^2 }}d\xi } . \\
 \end{array}
\end{equation}

The next task is to construct the expression of $\Phi$. One can verify that variable $\phi$ satisfies the Schwartzian form of variable coefficient SWW equation, so by substituting $\phi  = \frac{{\sqrt {c_4 \alpha } \tanh (\frac{{\sqrt {c_4 \alpha } (t + \Phi )}}{{m\beta }})}}{\alpha }$ into (\ref{eq4}), one can get,

\begin{equation}\label{eq16}
 \begin{array}{l}
 (2\tilde b m^2 \beta ^2  - 2mn\beta ^2 )\Phi _\xi  \Phi _\eta  \Phi _{\xi \xi }  + 4\rho \tilde a\alpha c_4 \Phi _\xi ^4 \Phi _{\xi \eta }  + ((2mn\beta ^2  - 2\tilde b m^2 \beta ^2 )\Phi _{\xi \eta }  - \rho \tilde a m^2 \beta ^2 \Phi _{\xi \xi \xi \eta } )\Phi _\xi ^2  \\
  + (\rho \tilde a m^2 \beta ^2 \Phi _{\xi \eta } \Phi _{\xi \xi \xi }  + 3m^2 \beta ^2 (2 + 3\rho \tilde a\Phi _{\xi \xi \eta } )\Phi _{\xi \xi } )\Phi _\xi   - 3\rho \tilde a m^2 \beta ^2 \Phi _{\xi \eta } \Phi _{\xi \xi }^2  = 0. \\
 \end{array}
\end{equation}

Eq.(\ref{eq16}) is still a partial differential equation. In order to get solutions to this equation, we do the traveling wave transform $\Phi(\xi,\eta)=\Phi(\xi+\mu\eta)=\Phi(\Delta)$ yields,

 \begin{equation}\label{eq17}
 \mu m^2 \beta ^2 \rho \tilde a \Phi _{\Delta \Delta \Delta \Delta } \Phi _\Delta ^2  - 4\rho \tilde a\mu \alpha c_4 \Phi _\Delta ^4 \Phi _{\Delta \Delta }  - m^2 \beta ^2 (2 + 4\rho \tilde a \mu \Phi _{\Delta \Delta \Delta } )\Phi _{\Delta \Delta } \Phi _\Delta   + 3\rho \tilde a\mu m^2 \beta ^2 \Phi _{\Delta \Delta }^3  = 0,
 \end{equation}
let $ \Phi _\Delta=G(\Delta)$, then,

 \begin{center}
 $\mu m^2 \beta ^2 \rho \tilde a G_{\Delta \Delta \Delta } G^2  - 4\rho \tilde a\mu \alpha c_4 G^4 G_\Delta   - m^2 \beta ^2 (2 + 4\rho \tilde a\mu G_{\Delta \Delta } )G_\Delta  G + 3\rho \tilde a\mu m^2 \beta ^2 G_\Delta ^3  = 0.$
 \end{center}

It is not difficult to verify that the above equation is equivalent to the following elliptic equation,

 \begin{equation}\label{eq18}
 G_\Delta   = \frac{{\sqrt {2\rho \tilde a\mu (2\rho \tilde a\mu \alpha c_4 G^4  - C_1 \rho \tilde a\mu m^2 \beta ^2 G^3  + C_2 \rho \tilde a\mu m^2 \beta ^2 G^2  + m^2 \beta ^2 G)} }}{{\rho \tilde a\mu m\beta }}.
 \end{equation}

It is know that Eq.(\ref{eq18}) has Jacobi elliptic functions solution. Hence, expression (\ref{eq14}) exhibit the interactions between soliton and abundant cnoidal periodic waves. We just take a simple solution of Eq.(\ref{eq18})  as

\begin{equation}\label{eq19}
G = a_0  + a_1 sn(\Delta ,l),
\end{equation}
where $a_0,a_1$ are undetermined constants.

By substituting Eq.(\ref{eq19}) into Eq.(\ref{eq18}) yields the following four solutions,

\begin{equation}\label{eq20}
\begin{array}{l}
 \{ a_0  = \frac{1}{{\rho \tilde a\mu (l^2  - 1)}},a_1  =  \pm a_0 ,\alpha  = \frac{{(m\beta \rho \tilde a\mu )^2 (l^4  - 2l^2  + 1)}}{{4c_4 }}\} , \\
 \{ a_0  =  - \frac{1}{{\rho \tilde a\mu (l^2  - 1)}},a_1  =  \pm a_0 ,\alpha  = \frac{{(m\beta \rho l \tilde a\mu )^2 (l^4  - 2l^2  + 1)}}{{4c_4 }}\} , \\
 \end{array}
\end{equation}
where $\rho,c_4,l,m,\beta,\tilde a$ and $\mu$ are arbitrary constants.

We can obtain expressions for $\Psi,U$ by substituting Eqs.(\ref{eq19},\ref{eq20}) with $ \Phi _\Delta=G(\Delta)$ into Eq.(\ref{eq15}). Then, the solutions to the variable coefficient SWW equation(\ref{eq1}) can be obtained accordingly by using (\ref{eq14}). Because the expression is too prolix,it is omitted here. Figure 1 displays a soliton-cnoidal wave solution, including the interaction of the kink soliton+cnoidal periodic wave of $u$ at $t = 0$. The parameters used in Figure 1 are selected as $\mu  = 2,l = 0.5,c_4  =  - 0.001,\rho  = 0.8,\tilde a = 0.1,m = 0.2,\beta  = 1$.


\section{Discussion and summary}

By developing the truncated Painlev\'{e} analysis, we investigate the B\"{a}cklund transformation and the Schwartzian form of the generalised variable coefficient shallow water wave equation. Nonlocal symmetries are also obtained by using a systematic method. The nonlocal symmetry group is revealed by solving the initial value problem of a closed system. In particular, by means of the symmetry reductions, several classes of new exact interaction solutions are provided in the paper. For example, general soliton-cnoidal wave solutions which can be explicitly expressed by the Jacobi elliptic functions are constructed. These results can be utilized to explain some important interaction phenomenons and more types of these soliton-cnoidal wave solutions need our further study.

\section{ Acknowledgments}

This work is supported by the National Natural Science Foundation of China (Nos. 11505090,1171041), the Natural Science Foundation of Shaanxi Province under Grant No.2018JQ-1045, and the Science and Technology Innovation Foundation of Xi¡¯an under Grant No.2017CGWL06.Research Award Foundation for Outstanding Young Scientists of Shandong Province(No. BS2015SF009). The doctorial foundation of Liaocheng University under Grant No.318051413.

\small{

\end{document}